\documentclass[pra,amsmath,amssymb,superscriptaddress,twocolumn]{revtex4}
\usepackage{graphicx}
\usepackage{color}
\newcommand{\halfsq}{\frac{1}{\sqrt{2}}}

\newcommand{\bla}{\color{black}}

\begin{document}

\title{The  wave-function  is  real   but  nonphysical:  A  view  from
  counterfactual  quantum  cryptography}  
\author{Akshata  Shenoy  H.}
\email{akshata@ece.iisc.ernet.in} 
\affiliation{Electrical  Communication Engg.  Dept.,  IISc, Bangalore,
  India}   
\author{R.    Srikanth}
\email{srik@poornaprajna.org}  \affiliation{PPISR,  Bangalore,  India}
\affiliation{Raman Research Institute, Bangalore, India}

\begin{abstract}
Counterfactual quantum  cryptography (CQC) is  used here as a  tool to
assess the status of the quantum state: Is it real/ontic (an objective
state of Nature)  or epistemic (a state of  the observer's knowledge)?
In contrast to  recent approaches to wave function  ontology, that are
based on realist models of quantum theory, here we recast the question
as  a  problem of  communication  between  a  sender (Bob),  who  uses
interaction-free measurements, and a receiver (Alice), who observes an
interference pattern  in a Mach-Zehnder  set-up.  An advantage  of our
approach is that  it allows us to define  the concept of ``physical'',
apart   from  ``real''.   In   instances  of   counterfactual  quantum
communication, reality is  ascribed to the interaction-freely measured
wave  function ($\psi$) because  Alice deterministically  infers Bob's
measurement.  On  the other  hand, $\psi$ does  not correspond  to the
physical transmission  of a particle because it  produced no detection
on Bob's  apparatus. We therefore  conclude that the wave  function in
this case  (and by  extension, generally) is  real, but  not physical.
Characteristically   for   classical   phenomena,  the   reality   and
physicality of objects are  equivalent, whereas for quantum phenomena,
the former is strictly weaker. As a concrete application of this idea,
the nonphysical reality  of the wavefunction is shown  to be the basic
nonclassical phenomenon that underlies the security of CQC.
\end{abstract}

\maketitle \date{}

\section{Introduction}

Quantum  superposition,  as exemplified  by  the familiar  double-slit
experiment, lies  at the heart  of quantum weirdness.   The experiment
contains,  according  to  Feynman,  the ``only  mystery''  in  quantum
mechanics   \cite{Fey63}.   The   interference  observed   behind  the
double-slit indicates that the  particle, when moving from the source,
travels in  some sense along \textit{both} slits.   Classical waves of
water or sound show such interfering behavior, but they are collective
phenomena,  and can  be  attributed to  distinct  localized masses  or
molecules travelling through the two  slits.  In the quantum case, the
interference can  be observed even  when the source is  so attentuated
that there  is practically only  one particle between  the double-slit
plane and the screen.  Dirac summarized the situation by saying ``Each
photon then interferes only with itself'' \cite{Dir30}.

Here we revisit  this phenomenon, prompted in part  by recent interest
in studying  the nature of the  quantum state.  The  basic, oldest and
arguably  most controversial question  here is:  Is the  quantum state
real, i.e., Does it objectively  exist and represent a state of Nature
(cf. eg., \cite{Kas12}); or, Is  it epistemic, i.e., Does it represent
the observer's  state of knowledge?   In the framework  of ontological
models of quantum mechanics \cite{HS10},  which is based on the hidden
variable theories of Bell \cite{Bel87} and Kochen-Specker \cite{KS67},
a  model  is  $\psi$-ontic  if  every  pair  of  pure  quantum  states
corresponds to non-overlapping ontic supports, and is $\psi$-epistemic
otherwise. Recently, a number of  arguments have been put forth within
this framework in favor of the ontic view \cite{PBR12,CR12,PPM13,CR13}
under certain assumptions \cite{LJB+12,ABC+13,GR0a,GR0b,GR0c,Lei14}.

Here  we will  assess the  status  of the  quantum state  from a  very
different  perspective,  one  that  differs  from the  above  type  of
approach to wave function ontology in three basic ways: (1) our result
is  not based  on the  framework  of realist  or any  other models  of
quantum  theory,  but  instead   one  that  is  based  on  operational
considerations   about    communication   involving   interaction-free
measurements  \cite{EV93}; (2)  The definition  of ``reality''  is not
explicitly tied to  assumptions like independence, no-signaling, etc.,
but is direct and intuitive; (3)  Our approach allows us to define not
only   a  concept   of  ``reality'',   but  also   another,   that  of
``physicality''.   We think  that this  is crucial,  because  our work
shows that the distinction between physicality and reality lies at the
heart of what makes the quantum state nonclassical.

\section{Interaction-free measurements and counterfactual
communication \label{sec:ifm}}

Interaction-free measurements  (IFMs) in  quantum theory allow  one to
obtain  knowledge of  the presence  or  absence of  an object  without
interrogating it directly.  For example,  an absorber placed in one of
arms of  a Mach-Zehnder interferometer is detected  by its disturbance
of  the  destructive interference  that  would  have  resulted in  its
absence \cite{EV93}.  IFM can be used as the basis for what are called
counterfactual   computation   \cite{Joz98,H+06}  and   counterfactual
crytpography \cite{N09,SSS1}, where the  absorber is present or absent
depending on  the outcome of a  computer or choice  of a communicator.
Here `counterfactual' essentially means  that Bob's blockade in one of
the  arms of  a Michelson  interferometer  causes a  detection of  the
particle away from the  blockade.  Single-particle nonlocality lies at
the heart of  this phenomenon, since without this,  the blockade would
localize a particle near it.

Recent  advances  include counterfactual  communication  based on  the
quantum Zeno  effect \cite{SLA+0} (cf.   also \cite{Vai13,SLA+14}) and
the  extension  of  the  counterfactual  bipartite  protocols  to  the
tripartite  case \cite{SSS14,Sal14}.  The  N09 protocol  of \cite{N09}
has  been experimentally implemented  \cite{BCD+12}, and  its security
proofs discussed by various authors \cite{YLC+10,ZWT+12,ZWT12}.

\begin{figure}
\includegraphics[width=6cm]{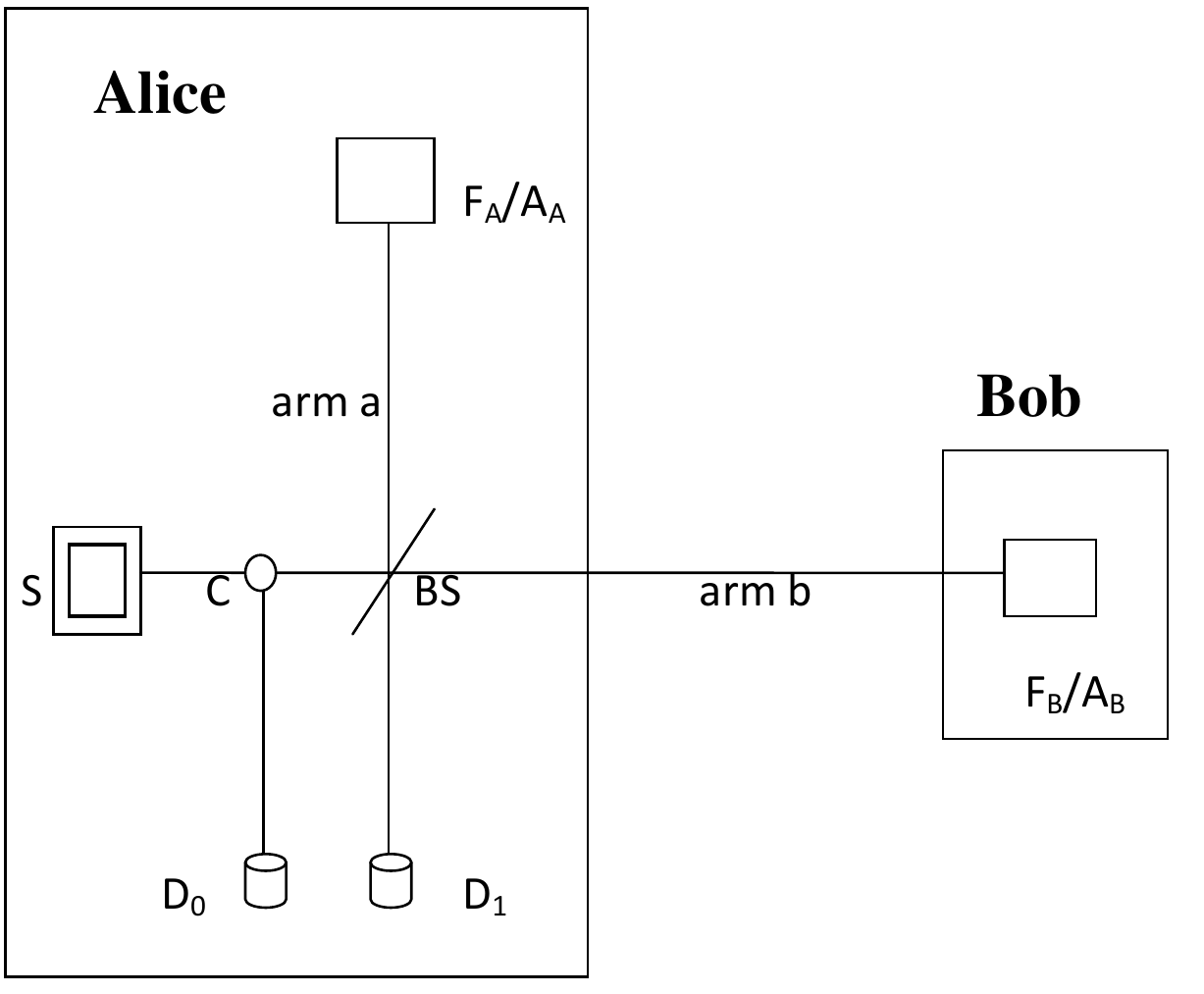}
\caption{Semi-counterfactual  quantum  cryptography:  Alice  inputs  a
  single-photon into  a beam-splitter  via an optical  circulator $C$,
  which  is split  along arm  $a$ (internal)  and $b$  (to  Bob). Each
  person  applies  either  operation  $F$ (reflect)  or  $A$  (absorb)
  randomly. If Alice  detects the photon in $D_0$,  she knows that her
  and  Bob's inputs  are anti-correlated.  Further, if  her  input was
  $F_A$, then  Bob's was $A_B$,  making the communication of  this bit
  counterfactual.}~\\ \hrule
\label{fig:cc_ontic}
\end{figure}

For the  purpose of this work, it  suffices for any one  outcome to be
counterfactual, and  we use the  recently proposed semi-counterfactual
quantum  cryptographic  scheme  \cite{SSS1},  where only  one  of  the
outcomes   is   counterfactual.     This   experimental   set-up   for
counterfactual   communication   is    based   on   a   Michelson-type
interferometer.  One of  the  arms is  taken  to be  an external  one,
connecting Alice and Bob, and the  other arm is an internal arm within
Alice's station (Figure \ref{fig:cc_ontic}).

A single-photon from a source in Alice's station hits her beamsplitter
BS and is  split along the two  arms $a$ and $b$.  Alice  and Bob each
possesses a switch which is randomly in one of two modes: absorb ($A$)
and reflect  with a Faraday mirror  ($F$).  The state  of the particle
after the beamsplitter is:
\begin{equation}
|\Psi\rangle             =             \left(\frac{a^\dag            +
  ib^\dag}{\sqrt{2}}\right)|0, 0\rangle_{AB},
\label{eq:Psi}
\end{equation}
where $a^\dag$ ($b^\dag$) is the  creation operator for the light mode
in the  internal (external) arm, and $|0,0\rangle_{AB}$  is the vacuum
state of those two modes. The operators for Alice's detector modes, in
terms of those for the arm modes, is given by:
\begin{equation}
d^\dag_j = \frac{1}{\sqrt{2}}\left(a^\dag + (-1)^j ib^\dag\right),
~(j=0,1)
\label{eq:trans}
\end{equation}
where   $d^\dag_j$  correspond   to  creation   operators   for  modes
corresponding  to  detector  $D_j$ (Figure  \ref{fig:cc_ontic}).   The
three possiblities of Alice's and Bob's choice are:
\begin{enumerate}
\item  Alice and Bob  both apply  $F$: This  results in an
  interference with a bright fringe at detector
  $D_1$ and a dark fringe at $D_0$:
\begin{equation}
P(D_1|F_A,F_B)=1.
\label{eq:D0FF}
\end{equation}
\item Exactly one of them applies $A$ and the other $F$: This produces
  a click in detector $D_0$ or $D_1$ each equal probability:
\begin{eqnarray}
P(D_0|A_A,F_B) &=& P(D_1|A_A,F_B)  = \frac{1}{2},\nonumber \\
P(D_0|F_A,A_B) &=& P(D_1|F_A,A_B) = \frac{1}{2},
\end{eqnarray}
where $A_A$  ($A_B$) indicates Alice  (Bob) applying $A$,  while $F_A$
($F_B$) indicates Alice (Bob) applying $F$.
\item Both apply $A$: neither detector $D_0$ nor $D_1$ clicks:
\begin{equation}
P(D_0|A_A,A_B)=P(D_1|A_A,A_B)=0.
\end{equation}
\end{enumerate}
Ideally, a bit is communicated  when Alice observes $D_0$, for in this
case   she  knows   for  certain   that  Bob   applied   an  operation
anti-correlated  with  hers.   The   efficiency  of  the  protocol  is
$\frac{1}{8}$  \cite{SSS1}.  

The counterfactual  situation arises in $D_0$ events  when Bob applies
$A_B$ and  Alice applies $F_A$.  In  this case, if  the detector $D_0$
clicks,  then it  carries 1-bit  of information  of  Bob's measurement
choice (that it was $A_B$),  even though the photon did not physically
travel and  interact with  his detector, as  evidenced by his  lack of
detection.   It  is in  this  sense  that  this 1-bit  information  is
counterfactual, and the  physical interpretation of this communication
forms the objective of this work.

The other  possibility for a $D_0$  event, that Bob  applied $F_B$ and
Alice  $A_A$, is  not counterfactual  with respect  to Bob  (hence the
characterization   of   the   protocol   of   Ref.    \cite{SSS1}   as
semi-counterfactual.)

\section{Physical interpretation} 

We distinguish two  concepts to be used to describe  the nature of the
quantum  state:  the  term  ``physical'', as  used  in  counterfactual
communication  to qualify  the  travel  of a  particle,  and the  term
``real'', as has  been used recently in quantum  foundations.  We will
define  these  two  concepts  based on  simple,  operational  criteria
inspired by the above experiment.

\subsection{Reality \label{sec:real}}

\begin{figure}
\includegraphics[width=8cm]{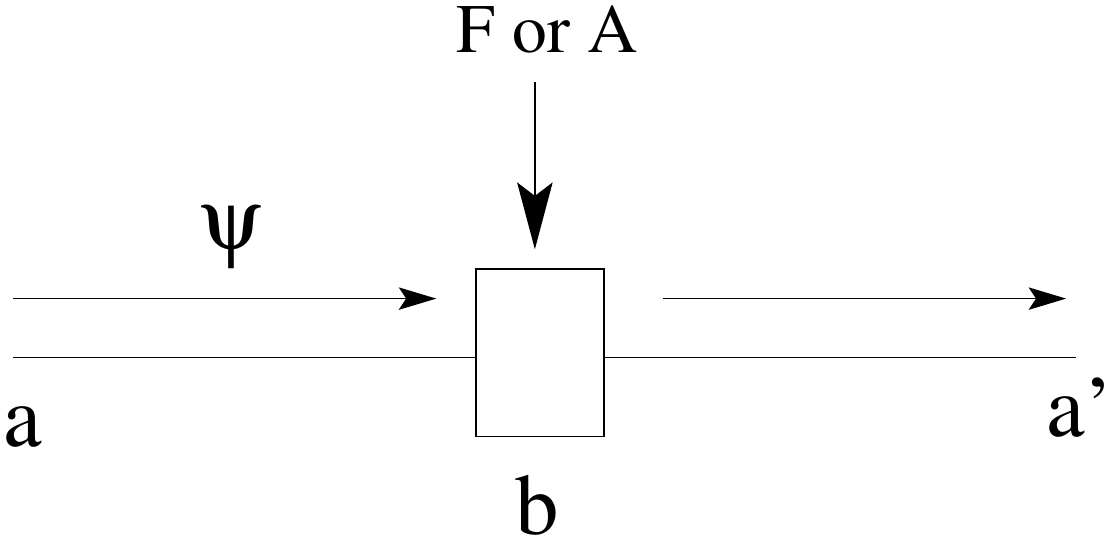}
\caption{Communication scenario  based on a spread out  version of the
  to-and-from particle travel along the external arm from Alice to Bob
  in Figure  \ref{fig:cc_ontic}: Alice knows that an  entity $\psi$ is
  transmitted from the  source located at $a$ towards  her, located at
  $a^\prime$.   Bob, located  at  the intermediate  position $b$,  may
  absorb  the  object  (operation   $A_B$)  or  forward  it  ($F_B$).}
~\\ \hrule
\label{fig:realphysical}
\end{figure}

Consider    the   communication    scenario   described    in   Figure
\ref{fig:realphysical}. This is an unwrapped version of the path taken
by a  particle from Alice's source,  via Bob, back to  Alice in Figure
\ref{fig:cc_ontic}.   Alice  knows that  a  certain  entity $\psi$  is
emitted  from the  source at  position  $a$ along  the indicated  path
towards position $a^\prime$. Bob's action at the intermediate position
$b$   is  to  block   or  to   not  block   its  passage   during  its
transit. Suppose $\psi$  is a real thing, like  a classical ball. Then
if Bob applies blocking action  $A_B$, Alice does not receive the ball
at $a^\prime$,  while if he  applies the forwarding action  $F_B$, she
does.  Thus  Bob's action can  be deterministically inferred  by Alice
through local observation.

On the other  hand, suppose $\psi$ is epistemic and  not a real thing.
Then it represents a probability distribution 
\begin{equation}
\textbf{P}_\psi \equiv (p_\psi, 1-p_\psi),
\label{eq:P}
\end{equation} 
as perceived  by Alice, with  $p_\psi$ being the probability  that the
particle   travels   from  $a$   towards   Alice   through  $b$,   and
$\overline{P}_\psi \equiv 1-p_\psi$, the  probability that it does not
travel from the source along this path towards Bob.

If Alice  finds the particle at  $a^\prime$, then Bob  clearly did not
apply $A_B$.   But if she does  not recover the particle,  it does not
necessarily  imply   that  Bob  applied   $A_B$,  since  there   is  a
nonvanishing probability $\overline{p}_\psi$ that the particle did not
travel  from  $a$  to  $b$.   In  other  words,  an  epistemic  $\psi$
necessarily  precludes  Alice's  deterministic retrodiction  of  Bob's
blocking action.

If  we denote  by  ``no''  Alice's non-detection  of  the particle  at
$a^\prime$, then $P(\textrm{no}|A_B)=1$, and by Bayesian argument:
\begin{equation}
P(A_B|\textrm{no}) = \frac{P(A_B)}{\overline{p}_\psi + p_\psi P(A_B)},
\label{eq:real}
\end{equation}
where  $P(A_B)$  is  the  probability  that Bob  applies  $A_B$.   Eq.
(\ref{eq:real})   shows  that   $P(A_B|\textrm{no})  =   1$   only  if
$\overline{p}_\psi=0$.   Otherwise,  $P(A_B|\textrm{no})<1$,  implying
that  Alice cannot  deterministically  infer that  Bob applied  action
$A_B$.  More  generally, ``yes'' indicates a specific  type of outcome
observed  by  Alice, conditioned  on  her  any  local operation.   For
example, if she turns on a knob, and a beep appears on a monitor, this
could  correspond to  ``yes'', while  ``no'' would  correspond  to the
non-appearance of such a beep.

This motivates the following definition of reality: We say that $\psi$
is  \textit{real} and  not epistemic,  if and  only if  there  is some
detection  event $\alpha^\prime$  at  $a^\prime$ such  that Alice  can
\textit{deterministically} infer  by local observation  in her station
that  Bob  applied  the  blocking  action $A_B$  during  the  $\psi$'s
transit:
\begin{equation}
\textbf{Real}(\psi)           \equiv           \exists_{\alpha^\prime}
       [P_{\alpha^\prime}(A_B|\textrm{no}) = 1],
\label{def:real}
\end{equation}
where  \textbf{Real}($\psi$)  is  the  proposition  that  asserts  the
reality  of $\psi$.   Note that  this condition  for reality  is quite
operational: it  is based on no interpretational  framework for $\psi$
but is stated in terms of  Bob's actions and Alice's observations in a
laboratory.

We now apply the above definition to a classical object.  If $\psi$ is
(say) a physical  ball transmitted along the indicated  path in Figure
(\ref{fig:realphysical}) with $p_\psi = 1$ in Eq.  (\ref{eq:P}).  Then
$\psi$  is  real  according   to  the  above  criterion,  since  every
non-detection  event  at  $a^\prime$   allows  Alice  to  infer  Bob's
absorbing action.   Further, if $\psi$ is  epistemic (corresponding to
$0  < p_\psi <  1$ in  Eq.  (\ref{eq:P})),  then it  fails to  be real
according to our criterion (\ref{def:real}).
  
Now  consider the  set-up in  Figure \ref{fig:cc_ontic}.   To  map the
scheme of  Figure \ref{fig:realphysical}, we let  both coordinates $a$
and  $a^\prime$ to  be with  Alice, and  $b$ with  Bob.  We  denote as
``yes'' the  $D_1$ events conditioned on Alice  applying the operation
$F$  locally.   Thus  the  observation  of $D_0$  conditioned  on  her
applying $F$ corresponds to  outcome ``no'' in Eq. (\ref{eq:real}). In
such cases, Bob's blocking  action $A_B$ is deterministically inferred
by Alice, since
\begin{eqnarray}
P(A_B|D_0,F_A) &=& 1 - P(F_B|D_0,F_A) \nonumber \\
  &=& 1 - \frac{P(D_0|F_A,F_B)P(F_B|F_A)}{P(D_0|F_A)} \nonumber \\
  &=&1,
\label{eq:detA}
\end{eqnarray}
by   Bayes'    theorem,   and   since    $P(D_0|F_A,F_B)=0$   by   Eq.
(\ref{eq:D0FF}),  while   $P(D_0|F_A)>0$.   According  to   the  above
criterion, then the wave function  $\psi_b$ that leaves the source and
travels to Bob, must be real. \bla

\subsection{Physicality \label{sec:physical}}

It is  in counterfactual quantum  communication that we  encounter the
idea  of ``transfer  of  information without  any  physical travel  of
particles'' or  ``non-physical transmission of  information''.  In the
experiment  of  Figure  \ref{fig:cc_ontic},  we  saw  that  it  arises
naturally from the way communication  occurs from Bob to Alice when he
measures the photon via interaction-free measurement, and she observes
$D_0$.

By  contrast, in  all foundational  approaches  to the  status of  the
wavefunction,  it is  only the  \textit{reality} of  the wavefunction,
rather than its \textit{physicality}, that is called into question. We
want to  stress that this  is the significant contribution  of quantum
cryptography,  in  particular  the  counterfactual  variety,  to  this
discussion.

To be  precise, the particle's  transmission is called  nonphysical in
counterfactual  cryptography  when  Bob's communication  generates  no
detection  on  his  apparatus.   Generalizing  this,  the  essence  of
physicality here, then,  is that a physical object  should be detected
by an ideal detector.

This motivates us to identify physicality with physical detectability:
We  say that  $\psi$ is  \textit{physical} if  and only  if  for every
transmission  event $\alpha$,  if  Bob applies  the absorption  action
$A_B$  during  $\psi$'s transit,  then  it  necessarily  leads to  his
detection (provided he has an  ideal detector), else (i.e., he applies
$F_B$), it leads to a detection  by Alice.  In other words, a physical
$\psi$ is always detected by an intercepting ideal detector:
\begin{equation}
\textbf{Physical}(\psi) \equiv \forall_\alpha 
[P_\alpha(\textrm{Y}|A_B) = 1,~ P_\alpha(\textrm{yes}|F_B) = 1],
\label{def:physical}
\end{equation}
where ``Y''  indicates Bob's  detector registering a  detection.  Here
\textbf{Physical}($\psi$) is the proposition which asserts that $\psi$
is physical.  This condition,  like Eq.  (\ref{def:real}) for reality,
is operational: it is not  based on any interpretational framework for
$\psi$,  but is  stated  in terms  of  Alice's and  Bob's actions  and
observations.

Note that  if $\psi$  is physical, then  from the second  condition in
(\ref{def:physical}),  we  find  that   $\psi$  is  also  real.   Thus
physicality entails  reality. On the  other hand, the converse  is not
true. $\textbf{Real}(\psi)$ only  entails that there are \textit{some}
events such that if Bob applies $F$, Alice finds ``yes''. If this does
not hold  true in  \textit{all} cases, then  it fails to  be physical,
opening up the possibility of nonphysical but real $\psi$.

In   Figure   \ref{fig:realphysical},   consider   $\psi$  to   be   a
\textit{classical} real  object.  Then $\textbf{P}_\psi  \equiv (1,0)$
and Bob's action  $A_B$ always leads to a  detection; while his action
$F_A$ leads to Alice's  observation of ``yes'', thereby satisfying the
physicality  condition   (\ref{def:physical}).   A  \textit{classical}
epistemic   object   fails   (\ref{def:physical})   because   in   the
probabilistic event when  no particle travels down to  Bob, his action
$A_B$ does not generate a  detection for him.  Thus, classical objects
are either real  and physical or epistemic and  nonphysical.  In other
words, reality and physicality are equivalent in this formalism.

Now consider the set-up  in Figure (\ref{fig:cc_ontic}).  In the $D_0$
events,  Bob's interaction-free  action  $A_B$ does  not generate  any
detection  on  his side,  so  that it  fails  the  first condition  in
(\ref{def:physical}), making it nonphysical.

\subsection{Nonphysical reality}

The defintions Eq. (\ref{def:real}) and (\ref{def:physical}) allow the
existence  of  objects that  are  real-physical, real-nonphysical  and
epistemic-nonphysical  in  the  present  framework.  As  noted  above,
classical  $\psi$ conflates  reality with  physicality. Thus  the only
classical  possibilities are  real-physical  and epistemic-unphysical.
Quantum mechanics allows  the additional existence of real-nonphysical
$\psi$.

Combining  the  results  from  the  Subsections  (\ref{sec:real})  and
(\ref{sec:physical}), we conclude that  the quantum wave function that
propagates from  Alice to Bob in  the onward leg is  real according to
criterion  (\ref{def:real})  but  nonphysical according  to  criterion
(\ref{def:physical}).  On the other hand, as noted above:
\begin{equation}
\textbf{Physical}(\psi) \Longrightarrow \textbf{Real}(\psi),
\label{eq:realphysical}
\end{equation}
so that  the class  of real  objects is strictly  weaker than  that of
physical  objects.   Thus  an   epistemic  (i.e.,  unreal)  object  is
necessarily nonphysical, according to our criterion.

\begin{figure}
\includegraphics[width=8cm]{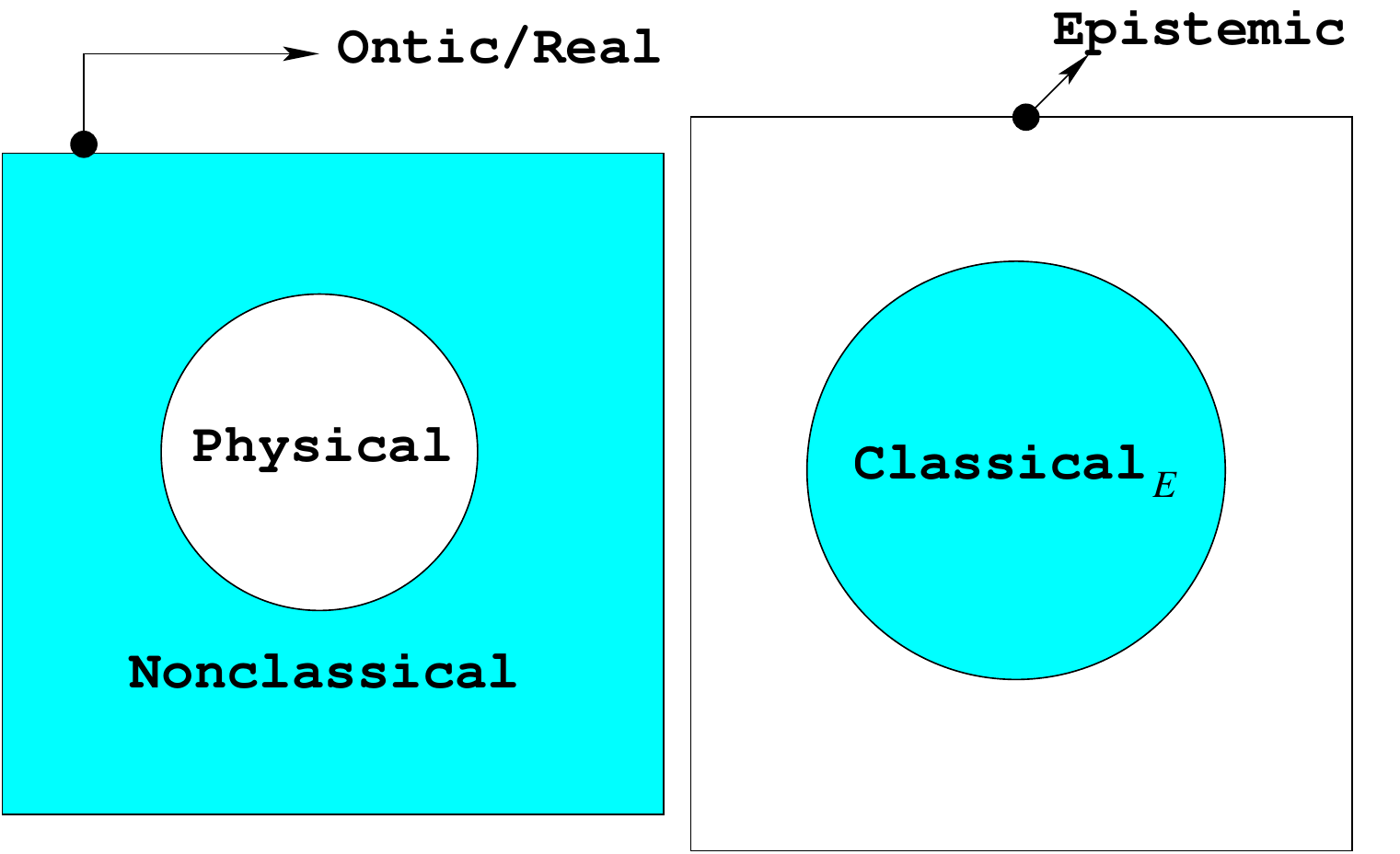}
\caption{The domain of real/ontic objects is strictly weaker than that
  of physical objects: $\texttt{Physical} \subset \texttt{Real}$.  The
  box  on  the  right   indicates  objects  obtained  via  probability
  distributions  over  objects in  the  box  on  the left  hand  side.
  Physicality  and  reality  are  equivalent  for  classical  objects:
  $\texttt{Classical} =  \texttt{Physical} \cup \texttt{Classical}_E$.
  Quantum  objects  comprise  all  objects above:  \texttt{Quantum}  =
  $\texttt{Real} \cup \texttt{Epistemic}$.}~\\
\hrule
\label{fig:nonclassical}
\end{figure}

Since  the equivalence  of reality  and physicality  holds  within the
domain of classical objects, objects that are real but nonphysical are
\textit{nonclassical}.    These   ideas   are   depicted   in   Figure
\ref{fig:nonclassical}, where  the set  of real (physical)  objects is
called \texttt{Real} (\texttt{Physical}).
\bla

Note  that classical  waves  (say water  waves)  show an  interference
behavior similar to the single-photon system. Yet water waves are both
real and physical: $\psi$ here  can be taken to represent the relative
mass of  water moving down each  arm, and ``yes''  can represent Alice
observing a depletion of water mass at her end, while $A_B$ represents
Bob's  blocking action.  It immediately  follows that  classical water
waves are physical, and  consequently real by criteria (\ref{def:real})
and (\ref{def:physical}).

\section{Security via nonphysical reality
of the wave function}

We now  claim that nonphysical  reality underlies the security  of the
semi-counterfactual cryptography protocol  of Section \ref{sec:ifm} in
the precise sense that the security check for the protocol consists in
verifying that the reality condition (\ref{def:real}) holds, while the
physicality (\ref{def:physical})  fails.  Suppose an  eavesdropper Eve
attacks the  exposed arm ($b$) in Figure  \ref{fig:cc_ontic}, with the
aim of testing whether the particle travelled along that path.  To the
extent  she  can  detect  the  particle's location,  she  commits  the
particle  to one  or the  other  arm, and  thereby enforces  classical
behavior.  The  degree of failure of Alice  to verify nonphysical-real
behavior in the data, alerts Alice and Bob to Eve's presence.

Let us consider this more quantitatively, a detailed study of which is
presented in  \cite{SSS1}.  Eve attacks the  particle in arm  $b$ at a
point along the solid line in  the onward leg, by first interacting it
with a probe of hers prepared in the initial state $|0\rangle_E$ using
interaction $V$:
\begin{equation}
\mathcal{V}|\Psi\rangle_{AB}|0\rangle_E             =             \halfsq(a^\dag
|0,0\rangle_{AB}|{-}\rangle_E + b^\dag|0,0\rangle_{AB}|{+}\rangle),
\label{eq:V}
\end{equation}
where $|{+}\rangle$ and $|{-}\rangle$ represent non-orthogonal states,
with $|\langle{-}|{+}\rangle|  = \cos\upsilon$, where  $\upsilon$ is a
security parameter  in the range  $[0,\pi/2]$.  Eve's intention  is to
subsequently  measure the probe,  and use  its outcome  in conjunction
with Alice's announcement, to optimize  her guess of the shared secret
bit.

If Alice announces $D_0$, then Eve measures her probe according to the
following positive  operator-valued measure (POVM)  \cite{nc00}, which
can be shown to be optimal:
\begin{eqnarray}
P_\pm &=& \frac{1}{1+\cos(\upsilon)}(1 - |{\mp}\rangle
\langle\mp|), \nonumber \\
P_{0} &=& 1 - P_+ - P_-,
\label{eq:yesno?}
\end{eqnarray}
where outcome  $P_\pm$ indicates  a conclusive outcome,  while outcome
$P_0$ is inconclusive.  From  (\ref{eq:yesno?}), it follows that on an
ensemble of $|{+}\rangle$ and  $|{-}\rangle$, the probability that Eve
obtains  a conclusive  answer is  $1 -  |\langle{+}|{-}\rangle| =  1 -
\cos(\upsilon)$.  Clearly, this is also Eve's information $I_E$ on the
(unprocessed) secret key.

Eve's  presence  is  therefore  revealed through  the  departure  from
perfect   interference,    as   observed   by    Alice,   which   from
Eqs. (\ref{eq:V}) and (\ref{eq:trans}) is
\begin{equation}
P(D_0|F_A,F_B)        =        \left|\left|\frac{|{+}\rangle_E       -
  |{-}\rangle_E}{\sqrt{2}}\right|\right|^2
= \frac{1}{2}(1    -
\cos(\upsilon)),
\label{eq:dep}
\end{equation}
implying  a   visibility  of  $V  =   \cos(\upsilon)$.   If  condition
(\ref{def:real})  is perfectly satisfied,  then, conditioned  on Alice
applying $F$, if Bob applies $F$, then a ``yes'' outcome, namely $D_1$
detection,  must arise.   To the  degree that  $P(D_0|F_A,F_B)>0$, the
condition  (\ref{def:real})   fails,  and  the   second  condition  in
(\ref{def:physical})  also  fails.    To  summarize,  security  arises
because  Eve's   attack  physicalizes,  and   thus  classicalizes  the
particle,  which is  detected  by  Alice and  Bob  who are  monitoring
nonphysical-real behavior.

To complete the  discussion, the error $\epsilon$ produced  in the key
is that Alice's and Bob's  inputs are not anti-correlated when a $D_0$
click happens \cite{SSS1}:
\begin{eqnarray}
\epsilon   &=&  P(F_A,F_B|D_0) \nonumber 
\\ &=&  \frac{P(D_0|F_A,F_B)P(F_A,F_B)}{P(D_0)}  
= \frac{1-V}{2-V},
\label{eq:epsilon}
\end{eqnarray}
since  $P(D_0)  =  \sum_y  P(D_0|y)P(y)$  where  $y  \in  \{(A_A,A_B),
(A_A,F_B),   (F_A,A_B),    (F_A,F_B)\}$,   where   $P(D_0|A_A,A_B)=0$,
$P(D_0|A_A,F_B)=P(D_0|F_A,A_B)=\frac{1}{4}$                         and
$P(A_A,A_B)=P(A_A,F_B)=P(F_A,A_B)=P(F_A,F_B)=\frac{1}{4}$.  The mutual
information  between Alice  and Bob  is $I_B  =  1-H(\epsilon)$, where
$H(\cdot)$ is  the Shannon binary entropy. The  condition for positive
key rate is that $I_B-I_E\ge 0$ \cite{CK78}, or
\begin{equation}
V \ge H\left(\frac{1-V}{2-V}\right),
\label{eq:mutu}
\end{equation}
which implies that  the error rate $\epsilon$ must  be less than about
21\% for  security, and  thence, for proof  of non-physicality  of the
wave function.  

With more powerful models of  Eve's attack, this error threshold would
be lowered.  However, the basic idea-- that Eve's attack physicalizes,
and thus classicalizes the particle-- should intuitively be the same.

\section{Conclusions and Discussions}

We have  used counterfactual quantum  cryptography (CQC) as a  tool to
study  the ontology  of the  quantum  wave function.   This stands  in
contrast to approaches based  on ontological theories, where a concept
of  ``reality'' for  $\psi$  is defined,  and  under certain  arguable
assumptions, arguments  in favor of  the quantum state being  real are
presented. Furthermore,  they do not appear to  throw sufficient light
on what exactly is weird about quantum mechanics.

In  the   present  approach,   we  gave  operational   definitions  of
``reality''  and of  ``physicality''  of $\psi$.   In particular,  the
reality of $\psi$  is inferred from the fact  of genuine communication
from  Bob  to Alice  during  counterfactual  communication, while  its
nonphysicality is  inferred from the  fact that Bob makes  no physical
detection of  the particle during that communication.   Thereby we are
able to draw  conclusions that go beyond earlier  approaches: (a) that
the quantum state is real; (b)  that it is not physical; (c) that this
nonphysical  reality  is the  mark  of  nonclassical phenomena,  since
reality and physicality coincide for classical phenomena.

Of  course,  the  status  of  the  wave function  (as  being  real  or
epistemic) does not depend on Bob's choice of $A_B$ or $F_B$. Nor does
it depend on whether Bob is located at the end of arm $a$ or $b$. What
may  conclude  is  that the  each  of  the  superposed states  in  Eq.
(\ref{eq:Psi}), $\psi_a  \equiv a^\dag|0,0\rangle$ and  $\psi_b \equiv
b^\dag|0,0\rangle$, is  by itself  real-nonphysical, and thus,  so too
the particle  state state $|\Psi\rangle  = \frac{1}{\sqrt{2}}(\psi_a +
\psi_b)$  in  Eq. (\ref{eq:Psi})  is  also  real-nonphysical.  We  may
therefore  conclude   that  the  quantum  state   is  quite  generally
real-nonphysical.  In retrospect, we may  reflect in this new light on
the  wisdom of Feynman's  observation with  regard to  the double-slit
experiment, mentioned in the  opening paragraph. Our approach suggests
that in the production of fringes in the double-slit experiment, there
is indeed  some ``real stuff'' travelling  down both slits,  but it is
not physical.  This explication thus puts  (or so we hope!)  a name on
the mystery alluded to by Feynman.

In counterfactual communication, Bob's choice has a nonlocal influence
on  Alice's local observation.  This manifestation  of single-particle
nonlocality is  consistent with relativity, since Alice  must wait for
the time required by Bob's particle to return, before deciding whether
he applied $A_B$ or $F_B$, and thus respects signal locality.

From   the  perspective   presented  here,   the  essential   mark  of
non-classical  behavior, and  the  `mystery' of  the  quantum, is  the
non-vanishing  gap between the  real and  the physical.   An analogous
situation has been reported also in the study of quantum correlations,
where  nonclassicality   is  identified  with  the   gap  between  the
communication  cost $C$  of nonlocal  correlations \textbf{P}  and the
signal $S$ accessible within \textbf{P} \cite{ASiqsa}. Now $C$ is just
the signal required to  simulate the correlations using resources from
a  deterministic hidden  variable  (DHV) theory.   Thus the  `mystery'
highlighted by this  latter gap is that the signal  that arises at the
ontological level cannot  be used for signaling at  the physical level
(cf. \cite{Spe05,Kak0}).

Our work showed that the  non-physical reality of the wave function is
not an abstruse philosophical notion, but has the concrete application
of being responsible for security in CQC.  Finally, we venture that it
is the lack of distinction in  the literature between the real and the
physical aspect  that is responsible for the  historical difficulty in
interpretting the physical significance  of the quantum state.  In the
discussion pertaining to the  double-slit experiment, at first one has
the intuitive feeling that there is something real traveling down both
slits.  One then  subconsciously  maps this  real  thing to  something
physical.  But  clearly the  possibility  of  the  quantum wave  as  a
physical  entity  is one  that  we  would  consciously reject.   Thus,
psychologically speaking, a  person thinking about quantum foundations
is  caught in the  perpetual dilemma  of deciding  whether or  not the
quantum state  is real. It is  our belief that our  work resolves this
dilemma.
 
\begin{acknowledgments}
RS acknowledges support from the DST project SR/S2/LOP-02/2012.
\end{acknowledgments}

\bibliography{axta}

\end{document}